\documentclass{article}

\usepackage{cleveref}
\usepackage{amsmath}
\usepackage{graphicx}
\usepackage{comment}
\usepackage{subfloat}
\usepackage{subfig}
\usepackage{geometry}
\usepackage{tabularx}
\usepackage{listings}
\usepackage{sidenotes}
\usepackage{amssymb}
\usepackage[figuresright]{rotating}

\begin{document}

\title{Advances in computer architecture}

\author{Irfan Uddin}

\author{Irfan Uddin\\
University of Amsterdam, The Netherlands\\
mirfanud@uva.nl}

\maketitle

\begin{abstract}

In the past, efforts were taken to improve the performance of a processor via
frequency scaling. However, industry has reached the limits of increasing the
frequency and therefore concurrent execution of instructions on multiple cores
seems the only possible option. It is not enough to provide concurrent
execution by the hardware, software also have to introduce concurrency in order
to exploit the parallelism.

\end{abstract}

\setcounter{tocdepth}{1}
\tableofcontents

\newpage

\section{Introduction}
\label{sn:introduction}

In 1965 Intel's co-founder Gorden Moore presented the now famous \emph{Moore's
law} stating that: \emph{The number of transistors on a chip will roughly
double every 18 months}. Later in 2000 the 18 months were changed to two years
but the law is still followed by electrical engineers. To quote Prof. Yale N.
Patt: \emph{Moore's law is not the law of physics, it is merely the psychology
    of electrical engineers what they can do with silicon. The law is alive
    only because it is used by industries to compete with each
other}~\cite{yale_patt}.

Computer architects have consensus that in order to increase the performance of
the processor will likely require more transistors on the chip, to support ever
increasing software needs. While Moore's law seems to enable the increase in
the number of transistors, it is not yet clear how these transistors can be
used to achieve the desired improvements. In this paper we summarize this
argument and promotes multi-core design as a promising approach. We present
some technological competition to the Microgrid.

The rest of the paper is organized as follows.
In Section~\ref{sn:performance_frequency_scaling} we describe the performance
improvement in architecture by frequency scaling. We explain the automatic
exploitation of ILP to increase performance
in Section~\ref{sn:parallel_instr_stream} and the use of concurrency in hardware to
improve performance in Section~\ref{sn:performance_concurrency_hardware}. We show the
explicit concurrency in software to achieve performance
in Section~\ref{sn:performance_concurrency_software}. We give the details of modern
many-core systems in Section~\ref{sn:modern_many-cores} and conclude the paper
in Section~\ref{sn:conclusion}.

\section{Performance via frequency scaling}
\label{sn:performance_frequency_scaling}

Since the advent of microprocessors, computer architects have always been in
the quest to achieve the best performance by increasing the frequency of a
single chip. To main focus is to achieve the lowest possible latency in the
execution of individual instructions and highest possible throughput in the
execution of the program. In order to achieve this improvement in performance
via frequency scaling, there are two possibilities; reduce the line width of
the technology and hence increase the switching speed of the transistors or
reduce the amount of work done in a cycle by pipelining instruction
execution.

\subsection{Power consumption}

In order to achieve a higher frequency in a given technology a higher voltage
is required which means more power is consumed. However, power consumption is
at odds with "Green computing"~\cite{Kurp:2008:GC:1400181.1400186}. Also as the
power consumption increases, all kinds of problems come up: atom migration in
gates, change of electronic properties, wear and tear due to thermal expansion
of the material itself, etc. Furthermore, devices are getting portable, and
high voltage means constant access to power which is simply not practical in
a portable environment.

\subsection{Dissipation of heat}

The higher power consumed by a transistor is dissipated from the chip in the
form of heat which must be removed from the chip to avoid damaging it, as
silicon simply can not take heat. Increasing the frequency of the transistor
will increase the quantity of heat, and decreasing the size of the transistor
(decreases the energy required to switch a transistor) will generally increase
the density of heat. It is expensive to provide cooling to a heated processor
and requires extra space to install fans. Portable devices with this much heat
can burn the body of the user. There is also the fact that batteries store
limited energy and hence we want to minimise the power consumption in order to
maximise lifetime.

\subsection{Delay in wires}

The frequency of transistors can easily be improved, but wires are slow and
wire delay becomes slower for a given amount of power dissipation. A wire is
passive and can be made fast but then it will require more energy. It is
required to look at the relative speed when driving a wire by a minimum sized
transistor. Therefore the transaction from one part of the chip to another part
is limited by wires i.e. the frequency of transistors is not limited by the
number of transistors that can be manufactured on a chip, but actually limited
by the number of gates reachable in a
cycle~\cite{Matzke:1997:PSS:619022.620798,Agarwal:2000:CRV:342001.339691}.
Increasing the frequency of the transistor means that there is a higher
communication delay than the computation delay. 

In the past years the frequency of the processor has increased at the rate of
75\% per year (not any more) and the frequency of the DRAM has increased at the
rate of just 7\% per year~\cite{McKee:2004:RMW:977091.977115}. Because of this
difference, the DRAM latency has become increasingly large compared to the
processor's cycle time which is very high from processor's point of view. This
divergence in latency is also known as "Memory
wall"~\cite{Wulf:1995:HMW:216585.216588}. To avoid the delay for DRAM access,
concurrency is the next logical step i.e. at the time the processor sends
request to DRAM and then has to wait for the request to complete, the processor
should be able to perform some other activities concurrent to the DRAM
access.

\section{Performance via automatic exploitation of ILP}
\label{sn:parallel_instr_stream}

An instruction stream has a combination of memory and computational operations.
When the memory operations are issued, there can be some independent
computational instructions that can be executed. But additional hardware is
required to find these independent instructions and schedule them so that no
data hazard occurs. This technique is called automatic exploitation of
Instruction Level Parallelism (ILP)~\cite{Wall:1991:LIP:106973.106991}; and a
number of approaches are tried in this direction such as pipelining,
out-of-order execution, branch prediction and cache prefetching.

\subsection{Pipelining}

The function of each instruction (i.e. fetch, decode, execute etc.) can be
broken down in multiple sub-functions. Like a conveyor belt in a factory line,
these sub-functions can be pipelined in parallel where multiple instructions
coming one after another. Pipelining have enabled the simultaneous execution of
multiple instructions. The actual execution time of the individual instruction
remain the same, but the throughput of several instructions is improved
typically by $x$ times where $x$ is the length of the pipeline. e.g. in Pentium
machines the throughput of instructions was increased up to 20 times. In
addition, pipelining also enabled the processor to have a higher clock
frequency because the operations executed at each pipeline stage is simpler and
therefore takes a shorter cycle time.

\subsection{Out-of-order execution}

The instruction stream has a limited number of independent instructions that
can be executed in parallel. Often, an instruction is dependent on a previous
operation (not only loads but also long latency operations like mul, floating
point operations etc.) and must therefore stall the pipeline until the load operation
completes. Out-of-order execution is used to allow instructions in the
pipeline to overtake those issued earlier in order to avoid stalling the
pipeline completely. However, hardware logic and energy is required for the
dependency analysis to determine if instructions can overtake each other. Also,
since instructions are executing out-of-order, a reorder buffer is required for
the in-order completion to provide the expected
results~\cite{Vandierendonck:2007:BOE:1242531.1242548}. Out-of-order execution
introduces additional dependencies e.g. Write-After-Read and Write-After-Write
which are resolved by register renaming i.e. allowing the hardware to use more
registers than will fit in the instruction format. However it increases the
size of the register file, lengthens the pipeline and therefore increases the
branch penalties~\cite{Jesshope03}.

\subsection{Branch prediction}

The number of instructions between branches is typically very small (average is
less than 6 instructions~\cite{Sechler:1995:DSL:203133.203135}). Branch
prediction is required to keep the pipeline full; while the branch and
instructions before it are executing, we can fetch and decode instructions
after the branch. If the branch was predicted accurately we can already have
the next instructions in the pipeline, however if the branch was not predicted
correctly it will result in a pipeline bubble or many pipeline bubbles in more
complex processors. In addition, the effect of instructions must also be
cancelled, which might involve the roll back of side effects. It means that a
huge number of cycles and energy are lost on computing something that did
not really participate in the required computation of the program. The issue
with branch prediction is, if there are multiple tasks/processes the branch
predictor has to predict branches not in only one but multiple programs
interleaved over time. To keep the high accuracy across heterogeneous codes,
the size of the branch predictor must grow. Because of multitasking branch
prediction is an expensive approach.

\subsection{Cache prefetching}

To avoid the delay of a cache miss, a processor can look at the memory access
pattern of the program and guess what cache line will be accessed next and
prefetch the line. If the guess is correct, the large delay required to access
the memory is avoided, however if it is not correct a large amount of energy
and memory bandwidth is wasted to fetch the data from the off-chip memory that
will not be used. Intel Itanium, Pentium 4 and Transmeta Crusoe are some
examples where cache prefetching is used.

\subsection{Discussion}

Traditional and super-scalar machines~\cite{Johnson:1989:SPD:891364} implement
most of the techniques described above in order to achieve the best performance
by the automatic exploitation of ILP. However, because of the additional
hardware the microprocessors are getting complex, energy inefficient and not
scalable~\cite{Cotofana:1998:DCI:945405.938236}. Next to super-scalar machines,
VLIW~\cite{Fisher:1983:VLI:1067651.801649} machines have been introduced which in
general are more scalable than super-scalar machines but have a complicated
design and a complex code generation process. In addition the binary code for a
VLIW machine can not be used on the same VLIW machine with a different issue
width i.e. binary compatibility is not achieved~\cite{Jesshope03}.

Computer architects have tried hard to improve performance of the programs
through implicit ILP without changing anything in the software. Because of
implicit ILP, the programmers were not required to have a detailed knowledge of
the hardware. The programmer used to wait to buy until a new more powerful machine is
available in the market and then the software magically got faster. However,
the industry is reaching the limits of increasing the frequency of processors
to execute more instructions in one cycle~\cite{bell.06.jpp}.

It seems like the free lunch is over~\cite{citeulike:6643735}, programmers have
to take responsibility of parallelization in order to get more performance
and therefore they need to get familiar with the architecture and concurrent
execution model. Some automatic parallelization techniques are taking the
approach of abstracting away the architecture
details~\cite{Grelck:2007:SOS:1248648.1248654}, but can not exploit the
architectural resources fully as programmers can do manually with the
knowledge of the architecture. The introduction of concurrency in software
engineering, increases the complexity of the software, decreases the
productivity of the engineer, but increases the performance of the software. 

\section{Performance via concurrency in hardware}
\label{sn:performance_concurrency_hardware}

The increased number of transistors on a chip has enabled computer architects
to provide concurrent execution of independent instructions on multiple
execution units. However, the implicit concurrency that can be extracted
automatically in hardware is limited~\cite{bell.06.jpp} i.e. parallelism is not
possible unless concurrency is introduced explicitly in programs.

\section{Performance via explicit concurrency in software}
\label{sn:performance_concurrency_software}

Instead of using a single instruction stream and trying to improve the
performance of the stream by implicit parallelism, we can have multiple
instruction streams which can execute concurrently. These streams are called
"threads" and provide an independent flow of control within a process. A
thread is also called a "light-weight" process, as it has its own stack, local
variables and program counter. Context switching in threads is much cheaper
than in processes. Industry has realized that writing parallel programs to exploit
concurrency of the hardware can not be avoided in future.

Introducing concurrency in programs is difficult, also debugging these programs
require a lot of effort from programmers. The order of execution is known but
the order of execution of threads is not known which makes the behavior of
multiple threads in a parallel program difficult to predict. The
synchronization between threads to ensure that shared state is accessed
atomically can make the concurrency difficult to define and can significantly
affect the performance of the program in multi-core systems. It is true that
threads in some form were quite commonly used in networking from 1980s e.g.
multiple connections to HTTP servers. In networking, threads represent
independent transaction and in applications, threads interact with each other in
producing results. Because of the inherent difficulty,
parallel programming never became a mainstream programming
paradigm~\cite{Talia:1997:PCS:256175.256193} in software engineering
community in the past.

Despite being difficult, parallel programing is the most desirable technology
in the current state-of-the-art multi-cores processors. The performance of an
application can be improved only from parallelization on contemporary hardware. A
number of programming libraries have been developed to provide concurrency
constructs to expose concurrency in programs. POSIX
threads~\cite{Garcia:2000:PTL:348120.348381}, OpenMP~\cite{openmp},
SystemC~\cite{SystemC} etc. are some of the libraries that provide constructs
for creation, termination, synchronization and communication between threads.

While parallelization is desirable, the management of threads in software by
the contemporary hardware is expensive. Typically 10-100 thousand cycles are consumed
in creation or synchronization of threads. There is also a cost of context
switching. Therefore fine-grained parallelism can not be achieved by explicit
concurrency in software. The next logical step is to introduce concurrency at
multiple levels; applications, operating systems and hardware. We need threads
in software, we need threads in hardware and we need the management of threads
in hardware. The concurrency at all levels will exploit the maximum possible
parallelism.

\section{Modern multi-core and many-core systems} \label{sn:modern_many-cores}

Since parallelization is the only practical solution in current technology and
the number of transistors on a single chip is
growing~\cite{Sodan:2010:PVM:1749398.1749434}, therefore we claim that in the
future there will be large number of cores on a single chip where programmers
have to put effort to write parallel programs with the knowledge of the
underlying architecture. In this section we describe some state-of-the-art
multi- and many- core systems. Some of these cores are available commercially
while others are mainly in the research domain and are comparable to the
Microgrid. There may exist some other many-core systems but these are not
discussed in this section.

\subsection{Nvidia's GPGPU}

Nvidia's GPGPU (General Purpose Graphical Processing Unit)~\cite{nvidia} makes
use of a GPU as a co-processor to accelerate the execution in CPUs for
general-purpose computing. The acceleration happens by offloading some of the
computationally intensive and time consuming portions of the program from CPU
to GPU. The rest of the application still runs on the CPU. Therefore it is also
known as "heterogeneous" or "hybrid" computing. A CPU generally consists of
two, four or eight cores, while the GPU consists of hundreds/thousands of
smaller cores. All these cores work together to crunch through the data in
the application. From the user's point of view, the application runs faster
because it is using the massively parallel processing power of the GPU to boost
performance. CUDA (Computer Unified Device Architecture) is a parallel
programming model for GPU. It enables dramatic increase in computing
performance by harnessing the power of GPUs.

\subsubsection*{Discussion}

GPU is actually a specialized accelerator that is connected to traditional
single- or multi-cores processor. Therefore a large amount of work is required
for the transaction of data between processor and GPUs. Programmers have to
divide the problem into coarse sub-problems that can be solved independently.
In addition the programmers have to explicitly manage the memory and
concurrency i.e. a complex model of the architecture is forced in the mind of
the programmer.

GPU architecture is based on the SIMD model and therefore can efficiently
execute SIMD based applications. The SIMD architectures are very inefficient in
branching, as each branch path must be executed sequentially. GPUs can achieve
a very high performance in embarrassingly parallel applications. However
applications with dense communication between threads (e.g. FFTs, compression
algorithms, sorting etc.) can not achieve a very high performance compared to
other multi-cores processors~\cite{karel2012}. GPUs become really slow in
functional calls, e.g. FFT is an embarrassingly parallel application, but it
does not scale very well on GPUs because of function calls from the outer loop.

\subsection{Sun/Oracle's UltraSPARC Tx}

Sun Microsystems have introduced a RISC architecture named SPARC (Scalable
Processor ARChitecture) in 1987. Oracle then bought Sun Microsystems and they together
introduced UltraSPARC T1 microprocessor (code name Niagara) in 2005. This
continued as a series of UltraSPARC T2, UltraSPARC T3 and in 2011 UltraSPARC
T4.

The UltraSPARC T4~\cite{UltraSparcT4} has a 16-stage integer pipeline, 11-stage
floating point pipeline and 2 issue width. It has a thread priority mechanism
where one thread can get preferential access to a core's hardware to give
increased performance. It has 4 processors on a single die and each processor
consists of 8 cores therefore a total of 32 cores are available on a single
chip. Each core has 8 hardware threads i.e. 64 hardware threads on a processor
and 256 hardware threads on the chip. Each core can switch between eight
threads using a modified LRU (Least Recently Used) algorithm for thread
selection. Each core is associated with 16KB of L1 I-cache and D-cache and
128KB of L2-cache. Eight cores share 4MB L3-cache and the DDR is 1TB. Total
transistors count is approximately 855 millions. The frequency of every core
can be changed in the range of 2.85 and 3.0 GHz. The technology used is 40nm
CMOS and the total die size is $403mm^2$.

The UltraSPARC T4 processor has increased single-thread performance, while
maintaining the high multi-thread throughput performance, therefore
single-threaded applications can have an efficient execution. It automatically
switches to single-thread mode when only a single thread is active,
dedicating all resources to that thread's execution. While software can
activate up to eight threads on each core at a time, hardware dynamically and
seamlessly allocates core resources such as instruction, data, L2-caches and
TLBs, as well as out-of-order execution resources such as the 128-entry
re-order buffer in the core. The cores provide sophisticated branch prediction
and have the features for prefetching instructions and data. 

\subsubsection*{Discussion}

The UltraSPARC is addressing internet servers or any large scale systems. It
generally addresses coarse-grained or embarrassingly parallel applications and
therefore disregard desktop computing and fine-grained parallelism. A library
is used to map software threads to hardware threads and it has an overhead of
creation and synchronization~\cite{T3-2011}. It is based on SMP model and has
an increased single thread performance. But it suffers from the scalability of
the bandwidth and power consumption of the interconnection between processor
and memory.

The UltraSPARC T4 cores are complex (16 stage pipeline) and have out-of-order
execution, branch prediction and cache prefetching, which are energy
in-efficient features. It has some inefficiency coming from the use of a huge
shared L2-cache which is necessary for server application (large number of
synchronizations around data) but has a cost in silicon i.e. heat, energy and
wiring complexity.

\subsection{Tilera's TILE64}

Tilera has introduced TILE64~\cite{tile64}, based on MIMD model. It has 64
cores on a chip using 90nm CMOS. These cores are fully functional, programmable
and each is capable of running its own operating system. A group of cores can
also be used to run as a symmetrical multi-processing operating system. Every
core has a frequency range of 600 to 900 MHz. The cores are in-order, three-way
VLIW issue width and implement a MIPS-derived VLIW instruction set. The
pipeline has many (more than 6) stages. Each core has 32 general-purpose
registers and three functional units: two integer arithmetic logic units and a
load-store unit. Every core has L1-cache and L2-cache as well as a distributed
L3-cache.

Tilera's architecture eliminates the on-chip bus interconnection by placing a
communications switch on each core and arranging them in a grid fashion on the
chip to create an efficient two-dimensional traffic system for packets. This
technology is named as intelligent mesh (iMesh). iMesh is similar to mesh
network used in Intel's SCC or NoC in embedded systems, with the innovation
that the flow of the messages in the mesh network can dynamically be adapted
based on the load of the network.

Tilera's Multicore Development Environment (MDE) provides
programming framework that allows developers to scale
their applications to large-scale multicore systems. It has enable the
standard tools such as gcc, gdb and oprofile, to be multi-core aware so that
the developer can easily debug, profile and optimize code running across dozens
of cores.

\subsubsection*{Discussion}

Tilera's architecture does not include hardware support for floating point
operation and therefore is not suitable for scientific computing. It mainly
targets embedded applications e.g. video encoding/decoding and network packet
processing. It is programmed in such a way that requires using registers for
communication by the programmer. Which means there is more responsibility on
the part of the programmer or compiler in creating threads. In a way programmers
need to understand the architecture in detail in order to program it. Also
the mapping of threads in the program to the hardware requires a software
library and therefore the cost of creation, synchronization and mapping of
software threads to hardware can not be avoided.

\subsection{Intel's SCC}

Intel's SCC (Single-chip Cloud Computer)~\cite{intelscc} is an experimental
many-cores research platform designed to address hardware and software
challenges by industry and academic institutions in the tera-scale
project~\cite{terascale}. It consists of 48 Pentium 1 cores connected in a
mesh network and on-chip message passing network for inter-thread
communication. The cores are relatively simple but fully functional
general-purpose cores. There is no hardware cache coherency protocol, which
allowed Intel to place 48 cores on a chip using CMOS 45nm technology. It does
not come as a stand-alone computer and a management PC (MCPC) is used to run the
applications on the chip.

Intel's SCC has fine-grained power management where software applications are
given control to turn cores on and off or to change their performance levels,
continuously adapting to use the minimum energy needed at a given moment. It
can run all 48 cores at one time over a range of 25W to 125W and selectively
vary the voltage and frequency of the mesh network as well as a group of cores.
Each tile (2 cores) can have its own frequency, and groupings of four tiles (8
cores) can each run at their own voltage. Every core uses the mainstream x86
(CISC) instruction set. The Linux operating system is available for the chip,
as well as gcc and Fortran compilers. A small library RCC is used for the
communication between cores.

\subsubsection*{Discussion}

Intel's SCC is a prototype which is designed for studying the parallel
programming paradigm in general-purpose computers. Therefore it is not really a
commercial product to be used for mainstream computing. It mainly addresses
coarse-grained parallelism, and may not achieve a high performance improvement
in fine-grained applications. The absence of hardware cache coherency protocol
places more responsibility on the programmer and hence requires more effort
from the programmers to manage the coherency of the caches. The Pentium 1 core
is actually single-threaded machine and therefore can not achieve
latency tolerance in long latency operations.

\subsection{Microgrid}

The
Microgrid~\cite{conf:hpc:Jesshope04,Bernard:2010:RPM:2031978.2031994,JesshopeAPC08}
is a general-purpose, many-core architecture developed at the University of
Amsterdam which implements hardware multi-threading using data flow scheduling
and a concurrency management protocol in hardware to create and synchronize
threads within and across the cores on chip. The suggested concurrent
programming model for this chip is based on fork-join constructs, where each
created thread can define further concurrency hierarchically. This model is
called the microthreading model and is also applicable to current multi-core
architectures using a library of the concurrency constructs called
\emph{svp-ptl} ~\cite{SVP-PTL2009} built on top of pThreads. In our work, we
focus on a specific implementation of the microthreaded architecture where each
core contains a single issue, in-order RISC pipeline with an ISA similar to
DEC/Alpha, and all cores are connected to an on-chip distributed memory
network~\cite{Jesshope:2009:ISM:1577129.1577136,Bousias:2009:IEM:1517865.1518255}.
Each core implements the concurrency constructs in its instruction set and is
able to support hundreds of threads and their contexts, called microthreads and
tens of families (i.e. ordered collections of identical microthreads)
simultaneously.

A number of tools and simulators are added to the designer's toolbox and used
for the evaluation of the Microgrid from different perspective. The compiler
for the Microgrid~\cite{poss.12.sl} can generate binary for different
implementations of the Microgrid. We have software libraries that provide the
run-time systems for the microthreading model on the shared memory SMP machines
and referred as \emph{svp-ptl}~\cite{SVP-PTL2009} and distributed memory for
clusters/grids and are referred as Hydra~\cite{Andrei:msc_hydra:2010} and
\emph{dsvp-ptl}~\cite{DSVP-PTL2011} The SL compiler can generate binary for
UTLEON3~\cite{5491777,danek.12},
MGSim~\cite{Bousias:2009:IEM:1517865.1518255,poss.13.MGSim.SAMOS} and
HLSim~\cite{Irfan:multipe_levels_hlsim:2013, Irfan:oneipc_hlsim:2013,
Irfan.12.2013.signatures, Irfan.12.2013.CacheBased, Irfan.01.2014.analytical,
Irfan:hl_sim_ptl:2011, Irfan:msc_hlsim:2009, Uddin:2012:CSM:2162131.2162132}.

\section{Conclusion}
\label{sn:conclusion}

The psychology of electrical engineers is that they can double the number of
transistors on a single chip every second year, which has enabled computer
architects to design more complex microprocessors. A number of approaches were
tried to achieve improvements in the throughput of the program implicitly. But
industry has reached the limits of implicit improvement in performance, and
multi-core designs seem to be promising approaches to achieve performance
explicitly. However, concurrency in hardware alone can not improve the
performance of the program unless concurrency is also exposed in software.





\bibliographystyle{plain}
\bibliography{main}







\end{document}